\documentstyle[12pt,epsfig]{article}
\begin{document}
\parskip=5pt plus 1pt minus 1pt

\begin{flushright}
{\bf hep-ph/0002246} \\
{\bf LMU-00-02}
\end{flushright}

\vspace{0.2cm}

\begin{center}
{\bf New Formulation of Matter Effects on Neutrino Mixing and 
$CP$ Violation}
\end{center}

\begin{center}
{\bf Zhi-zhong Xing} \footnote{Electronic address: 
xing@theorie.physik.uni-muenchen.de} \\  
{\it Sektion Physik, Universit$\it\ddot{a}$t M$\it\ddot{u}$nchen,
Theresienstrasse 37A, 80333 Munich, Germany}
\end{center}

\vspace{3cm}

\begin{abstract}
Within the framework of three lepton families we have derived an
exact and compact formula to describe the matter effect on
neutrino mixing and $CP$ violation. This model- and
parametrization-independent result can be particularly useful 
for recasting the fundamental lepton flavor mixing matrix from
a variety of long-baseline neutrino experiments.
\end{abstract}

\newpage
Recent observation of the atmospheric and solar neutrino anomalies
by the Super-Kamiokande Collaboration has provided robust 
evidence for massive neutrinos that mix among different flavors \cite{SK}
\footnote{Throughout this work the LSND 
evidence for neutrino oscillations \cite{LSND}, which has not been 
independently confirmed by other experiments \cite{KARMEN}, will not be
taken into account.}.
It opens a convincing window for new physics beyond the Standard Model, 
and has important cosmological implications.

Although the non-accelerator neutrino experiments have yielded
some impressive constraints on the parameter space of atmospheric and solar
neutrino oscillations, a precise determination of the lepton flavor
mixing angles and the leptonic $CP$-violating phase(s) has to rely on
a new generation of accelerator experiments with very long 
baselines \cite{Long}, including the 
possible neutrino factories \cite{Factory}. In such long-baseline
neutrino experiments the earth-induced matter effects, which are likely
to deform the neutrino oscillation behaviors in vacuum and
even to fake the genuine $CP$-violating asymmetries, must be
taken into account. To single out the ``true'' theory of lepton mass
generation and $CP$ violation depends crucially upon how accurately
the fundamental parameters of lepton flavor mixing can be measured
and disentangled from the matter effects. It is therefore desirable 
to explore the {\it analytical} relationship between the genuine flavor
mixing matrix and the matter-corrected one beyond the conventional
two-flavor framework \cite{Wolfenstein,Barger80,Langacker,Smirnov}. 
Some attempts of this nature have so far
been made \cite{Barger80,Zaglauer,Kuo,Barger99}, but they are 
subject to specific assumptions, approximations or parametrizations 
and need be substantially improved.

In this note we present an exact and compact formula to describe the
matter effect on lepton flavor mixing and $CP$ violation within the framework
of three lepton families. The result is completely independent of the
specific models of neutrino masses and the specific parametrizations of
neutrino mixing. Therefore it will be particularly useful, in the long
run, to recast the fundamental flavor mixing matrix from the precise
measurements of neutrino oscillations in a variety of long-baseline
neutrino experiments.

First let us define the $3\times 3$ lepton flavor mixing matrix
in vacuum to be $V$. It links the neutrino mass eigenstates 
($\nu_1, \nu_2, \nu_3$) to the neutrino flavor eigenstates 
($\nu_e, \nu_\mu, \nu_\tau$):
\begin{equation}
\left ( \matrix{
\nu_e \cr
\nu_\mu \cr
\nu_\tau \cr} \right ) =
\left ( \matrix{
V_{e1}  & V_{e2}        & V_{e3} \cr
V_{\mu 1}       & V_{\mu 2}     & V_{\mu 3} \cr
V_{\tau 1}      & V_{\tau 2}    & V_{\tau 3} \cr} \right )
\left ( \matrix{
\nu_1 \cr
\nu_2 \cr
\nu_3 \cr} \right ) .
\end{equation}
If neutrinos are massive Dirac fermions, $V$ can be parametrized
in terms of three rotation angles and one $CP$-violating phase.
If neutrinos are Majorana fermions, however, two additional
$CP$-violating phases are in general needed to fully parametrize $V$.
The strength of $CP$ violation in neutrino oscillations, no
matter whether neutrinos are of the Dirac or Majorana type, 
depends only upon a universal parameter $\cal J$ \cite{Jarlskog}:
\begin{equation}
{\rm Im} \left (V_{\alpha i}V_{\beta j} V^*_{\alpha j}V^*_{\beta i} \right )
= {\cal J} \sum_{\gamma,k} \epsilon^{~}_{\alpha \beta \gamma} 
\epsilon^{~}_{ijk} \; ,
\end{equation}
where $(\alpha, \beta, \gamma)$ and $(i, j, k)$ run over
$(e, \mu, \tau)$ and $(1, 2, 3)$, respectively. In the specific models 
of fermion mass generation $V$ can be derived from the mass matrices 
of charged leptons and neutrinos \cite{FX99}.
To test such theoretical models one has to compare their predictions
for $V$ with the experimental data of neutrino oscillations. The latter 
may in most cases be involved in the potential matter effects and must 
be carefully handled. 

In matter the effective Hamiltonian for neutrinos can be written 
as \cite{Wolfenstein}
\begin{equation}
{\cal H}_\nu = \frac{1}{2E} \left [ V \left (\matrix{
m^2_1     & 0      & 0 \cr
0         & m^2_2  & 0 \cr
0         & 0      & m^2_3 \cr} \right ) V^{\dagger}
+ \left ( \matrix{
A     & 0     & 0 \cr
0     & 0     & 0 \cr
0     & 0     & 0 \cr} \right ) \right ] ,
\end{equation}
where $m_i$ (for $i=1, 2, 3$) denote neutrino masses,
$A = 2\sqrt{2} G_{\rm F} N_e E$ describes the charged-current 
contribution to the $\nu_e e^-$ forward scattering,
$N_e$ is the background density of electrons, and $E$ stands for
the neutrino beam energy. The neutral-current contributions, which 
are universal for $\nu_e$, $\nu_\mu$ and $\nu_\tau$ neutrinos,
lead only to an overall unobservable phase and have been neglected.
Transforming ${\cal H}_\nu$ from the flavor eigenbasis to the
mass eigenbasis, we obtain the effective mass-squared
matrix of neutrinos:
\begin{equation}
\Omega_\nu = \left ( \matrix{
m^2_1 + A |V_{e1}|^2 &
A V^*_{e1} V_{e2} &
A V^*_{e1} V_{e3} \cr
A V_{e1} V^*_{e2} &
m^2_2 + A |V_{e2}|^2 &
A V^*_{e2} V_{e3} \cr
A V_{e1} V^*_{e3} &
A V_{e2} V^*_{e3} &
m^2_3 + A |V_{e3}|^2 \cr} \right ) ,
\end{equation}
This Hermitian matrix can be diagonalized through a unitary
transformation: $U^\dagger \Omega_\nu U = 
{\rm Diag} \{ \lambda_1, \lambda_2, \lambda_3 \}$,
where $\lambda_i$ denote the effective mass-squared eigenvalues of
neutrinos. Explicitly one finds
\footnote{The expressions of $\lambda_i$ in Ref. \cite{Barger80}
have apparent printing errors, while those given in Ref. \cite{Zaglauer}
depend upon a specific parametrization of $V$.}
\begin{eqnarray}
\lambda_1 & = & m^2_1 + \frac{1}{3} x - \frac{1}{3} \sqrt{x^2 - 3y}
\left [z + \sqrt{3 \left (1-z^2 \right )} \right ] , 
\nonumber \\ 
\lambda_2 & = & m^2_1 + \frac{1}{3} x - \frac{1}{3} \sqrt{x^2 -3y}
\left [z - \sqrt{3 \left (1-z^2 \right )} \right ] , 
\nonumber \\ 
\lambda_3 & = & m^2_1 + \frac{1}{3} x + \frac{2}{3} z \sqrt{x^2 - 3y} \; ,
\end{eqnarray}
where $x$, $y$ and $z$ are given by 
\begin{eqnarray}
x & = & \Delta m^2_{21} + \Delta m^2_{31} + A \; , 
\nonumber \\
y & = & \Delta m^2_{21} \Delta m^2_{31} + A \left [ 
\Delta m^2_{21} \left ( 1 - |V_{e2}|^2 \right ) 
+ \Delta m^2_{31} \left ( 1 - |V_{e3}|^2 \right ) \right ] , 
\nonumber \\
z & = & \cos \left [ \frac{1}{3} \arccos \frac{2x^3 -9xy + 27
A \Delta m^2_{21} \Delta m^2_{31} |V_{e1}|^2}
{2 \left (x^2 - 3y \right )^{3/2}} \right ] 
\end{eqnarray}
with $\Delta m^2_{21} \equiv m^2_2 - m^2_1$ and
$\Delta m^2_{31} \equiv m^2_3 - m^2_1$. 
Note that the columns of $U$ are the eigenvectors of $\Omega_\nu$ in the
vacuum mass eigenbasis. After a lengthy calculation, we arrive at
the elements of $U$ as follows:
\begin{equation}
U_{ii} = \frac{N_i}{D_i} , ~~~~~~~
U_{ij} = \frac{A}{D_j} \left (\lambda_j - m^2_k \right )
V^*_{ei} V_{ej} \; ,
\end{equation}
where $(i,j,k)$ run over $(1,2,3)$ with $i\neq j\neq k$, and
\begin{eqnarray}
N_i & = & \left (\lambda_i - m^2_j \right ) \left (\lambda_i - m^2_k \right )
- A \left [\left (\lambda_i - m^2_j \right ) |V_{e k}|^2  
+ \left (\lambda_i - m^2_k \right ) |V_{e j}|^2 \right ] ,
\nonumber \\
D^2_i & = & N^2_i + A^2 |V_{e i}|^2 \left [ 
\left (\lambda_i - m^2_j \right )^2 |V_{e k}|^2  
+ \left (\lambda_i - m^2_k \right )^2 |V_{e j}|^2 \right ] . 
\end{eqnarray}
One may check that $A=0$ leads to $\lambda_i = m^2_i$ and $D^2_i = N^2_i$,
and then $U$ becomes the unity matrix. 

It should be noted that the unitary matrix $U$ transforms the neutrino 
mass eigenstates in matter to those in vacuum. Therefore the lepton flavor 
mixing matrix in matter, denoted by $V^{\rm m}$, is a product of the lepton 
flavor mixing matrix in vacuum ($V$) and the matter-to-vacuum transformation 
matrix $U$: $V^{\rm m} = V U$. The elements of $V^{\rm m}$ turn out to be
\begin{eqnarray}
V^{\rm m}_{\alpha i} & = & \frac{N_i}{D_i} V_{\alpha i} 
+ \frac{A}{D_i} V_{e i} \left [ \left (\lambda_i - m^2_j \right )
V^*_{e k} V_{\alpha k} 
+ \left (\lambda_i - m^2_k \right )
V^*_{e j} V_{\alpha j} \right ] ,
\end{eqnarray}
where $\alpha$ runs over $(e, \mu, \tau)$ and $(i, j, k)$ over $(1, 2, 3)$
with $i \neq j \neq k$. Obviously $A=0$ leads to 
$V^{\rm m}_{\alpha i} = V_{\alpha i}$. This exact and compact formula shows 
clearly how the flavor mixing matrix in vacuum is corrected by the matter 
effects. Instructive analytical approximations can be made for Eq. (9),
once the hierarchy of neutrino masses is experimentally known or
theoretically assumed.

The result obtained above is valid for neutrinos interacting with matter.
As for antineutrinos, the corresponding formula for the flavor mixing 
matrix in matter can straightforwardly be 
obtained from Eq. (9) through the replacements 
$V \Longrightarrow V^*$ and $A \Longrightarrow -A$.

With the help of Eq. (9) we now calculate the universal
$CP$-violating parameter in matter, defined as ${\cal J}_{\rm m}$ through
\begin{equation}
{\rm Im} \left (V^{\rm m}_{\alpha i}V^{\rm m}_{\beta j} 
V^{\rm m *}_{\alpha j}V^{\rm m *}_{\beta i} \right )
= {\cal J}_{\rm m} \sum_{\gamma,k} \epsilon^{~}_{\alpha \beta \gamma} 
\epsilon^{~}_{ijk} \; ,
\end{equation}
where $(\alpha, \beta, \gamma)$ and $(i, j, k)$ run over
$(e, \mu, \tau)$ and $(1, 2, 3)$, respectively. The unitarity
of $V^{\rm m}$ allows us to simplify the lengthy calculation
and arrive at an instructive relationship between ${\cal J}_{\rm m}$ 
and ${\cal J}$:
\begin{equation}
{\cal J}_{\rm m} = {\cal J} \frac{\Delta m^2_{21} ~ \Delta m^2_{31}
~ \Delta m^2_{32}}{(\lambda_2 - \lambda_1) ~ (\lambda_3 - \lambda_1)
~ (\lambda_3 - \lambda_2 )} \; \; .
\end{equation}
The same result has been obtained by Harrison and Scott using the
Jarlskog determinant of lepton mass matrices, which is invariant 
for neutrinos in vacuum and in matter \cite{Harrison}.
Eq. (11) indicates that the matter contamination to 
$CP$- and $T$-violating
observables is in general unavoidable. However, $T$ violation
is expected to be less sensitive to matter effects than $CP$ violation, 
since the former is associated only with either neutrinos ($+A$)
or antineutrinos ($-A$) while the latter is related to both of them.

To be more explicit we calculate the conversion probabilities of 
$\nu_\alpha$ (or $\bar{\nu}_\alpha$) 
to $\nu^{~}_\beta$ (or $\bar{\nu}^{~}_\beta$) neutrinos in 
matter. We obtain
\begin{eqnarray}
P_{\rm m}(\nu_\alpha \rightarrow \nu^{~}_\beta) & = &
-4 \sum_{i<j} [ {\rm Re} ( V^{\rm m}_{\alpha i} V^{\rm m}_{\beta j} 
V^{\rm m *}_{\alpha j} V^{\rm m *}_{\beta i} ) 
~ \sin^2 \Delta_{ij} ] 
+ 8 {\cal J}_{\rm m} \prod_{i<j} \sin \Delta_{ij} \; ,
\nonumber \\
P_{\rm m}(\bar{\nu}_\alpha \rightarrow \bar{\nu}^{~}_\beta) & = &
-4 \sum_{i<j} 
[ {\rm Re} ( \tilde{V}^{\rm m}_{\alpha i} \tilde{V}^{\rm m}_{\beta j} 
\tilde{V}^{\rm m *}_{\alpha j} \tilde{V}^{\rm m *}_{\beta i} ) 
~ \sin^2 \tilde{\Delta}_{ij} ] 
- 8 \tilde{\cal J}_{\rm m} \prod_{i<j} \sin 
\tilde{\Delta}_{ij} \; ,
\end{eqnarray}
where $(\alpha, \beta)$ run over $(e, \mu)$, $(\mu, \tau)$ or $(\tau, e)$;
$\tilde{V}_{\alpha i}(A) \equiv V_{\alpha i}(-A)$,
$\tilde{\Delta}_{ij}(A) \equiv \Delta_{ij}(-A)$, and $\tilde{\cal J}_{\rm m}(A)
\equiv {\cal J}_{\rm m}(-A)$; 
and $\Delta_{ij} \equiv 1.27 (\lambda_i - \lambda_j)L/E$ with $L$ the
distance between the production and interaction points of $\nu_\alpha$
(in unit of km) and $E$ the neutrino beam energy (in unit of GeV). 
The probabilities of $\nu^{~}_\beta \rightarrow \nu_\alpha$ and
$\bar{\nu}^{~}_\beta \rightarrow \bar{\nu}_\alpha$ transitions
can be read off from Eq. (11) with the replacements ${\cal J}_{\rm m}
\Longrightarrow -{\cal J}_{\rm m}$ and $\tilde{\cal J}_{\rm m}
\Longrightarrow -\tilde{\cal J}_{\rm m}$, respectively. 

One can then define the
$CP$- and $T$-violating asymmetries as
\begin{eqnarray}
{\cal A}_{CP} & = & \frac{P_{\rm m}(\nu_\alpha \rightarrow \nu^{~}_\beta)
- P_{\rm m}(\bar{\nu}_\alpha \rightarrow \bar{\nu}^{~}_\beta)}
{P_{\rm m}(\nu_\alpha \rightarrow \nu^{~}_\beta)
+ P_{\rm m}(\bar{\nu}_\alpha \rightarrow \bar{\nu}^{~}_\beta)} \; ,
\nonumber \\
{\cal A}_{T} & = & \frac{P_{\rm m}(\nu_\alpha \rightarrow \nu^{~}_\beta)
- P_{\rm m}(\nu^{~}_\beta \rightarrow \nu_\alpha)}
{P_{\rm m}(\nu_\alpha \rightarrow \nu^{~}_\beta)
+ P_{\rm m}(\nu^{~}_\beta \rightarrow \nu_\alpha)} \; .
\end{eqnarray}
Note that ${\cal A}_T = {\cal A}_{CP}$ holds in vacuum (i.e., $A=0$),
as a consequence of $CPT$ invariance. Any
discrepancy between these two observables will definitely measure 
the matter effects in long-baseline neutrino experiments.

Finally let us give a numerical illustration of the matter-induced 
corrections to the flavor mixing matrix and $CP$ (or $T$) violation in vacuum.
The elements of $V^{\rm m}$, except the Majorana phases,
can be completely determined by four rephasing-invariant quantities
(e.g., four independent $|V^{\rm m}_{\alpha i}|$ or three independent 
$|V^{\rm m}_{\alpha i}|$ plus ${\cal J}_{\rm m}$). As the solar
and atmospheric neutrino oscillations in vacuum are essentially associated with
the elements in the first row and the third column of $V$,
it is favored to choose $|V_{e1}|$, $|V_{e2}|$, 
$|V_{\mu 3}|$ and ${\cal J}$ as the four basic parameters. 
For illustration we 
take $|V_{e1}| = 0.816$, $|V_{e2}| = 0.571$, $|V_{\mu 3}| = 0.640$,
and ${\cal J} = \pm 0.020$ for neutrinos and antineutrinos
\footnote{This specific choice corresponds to $\theta_{12} \approx 35^\circ$,
$\theta_{23} \approx 40^\circ$, $\theta_{13} \approx 5^\circ$, and
$\delta \approx \pm 90^\circ$ in the PDG-advocated parametrization of $V$
\cite{PDG98}.}.
Such a choice is consistent with the CHOOZ experiment \cite{CHOOZ}, the
large-angle MSW solution to the solar neutrino problem \cite{Bahcall}, 
and a nearly maximal mixing in the atmospheric neutrino oscillation. 
The relevant neutrino mass-squared differences are typically taken to be
$\Delta m^2_{21} = 5 \cdot 10^{-5} ~ {\rm eV}^2$ and
$\Delta m^2_{31} = 3 \cdot 10^{-3} ~ {\rm eV}^2$ \cite{SK,Bahcall}.
To calculate the $CP$- and $T$-violating asymmetries in the long-baseline
neutrino experiments, we assume a constant earth density profile and take 
$A = 2.28 \cdot 10^{-4} ~ {\rm eV}^2 E/[{\rm GeV}]$ \cite{Barger99}. We also
choose the baseline length $L = 732$ km, corresponding to a neutrino
source at Fermilab pointing toward the Soudan mine or that at
CERN toward the Gran Sasso underground laboratory. Using these
inputs as well as Eqs. (9)--(12), we first take $(\alpha, \beta) = (e, \mu)$
and compute the 
asymmetries ${\cal A}_{CP}$ and ${\cal A}_T$ changing with $E$
in the range $2 ~ {\rm GeV} \leq E \leq 30 ~ {\rm GeV}$. Then we
compute the ratios $|V^{\rm m}_{\alpha i}|/|V_{\alpha i}|$ and
${\cal J}_{\rm m}/{\cal J}$ as functions of the matter parameter $A$,
instead of $E$, in the range 
$10^{-7} ~ {\rm eV}^2 \leq A \leq 10^{-2} ~ {\rm eV}^2$. The 
numerical results are shown in Figs. 1 and 2.

We observe that matter effects can be significant for the elements in 
the first and the second columns of $V$, if $A \geq 10^{-5} ~ {\rm eV}^2$. 
In comparison, the magnitudes of $|V_{e3}|$, $|V_{\mu 3}|$ and
$|V_{\tau 3}|$ may be drastically enhanced or suppressed only for
$A > 10^{-3} ~ {\rm eV}^2$. The neutrinos are relatively more sensitive to
the matter effects than the antineutrinos.

The magnitude of ${\cal J}_{\rm m}$ decreases, when the matter
effect becomes significant (e.g., $A \geq 10^{-4} ~{\rm eV}^2$).
However, this does not imply that the $CP$- or $T$-violating asymmetries 
in realistic long-baseline neutrino oscillations would be smaller than
their values in vacuum. Large matter effects can significantly modify  
the frequencies of neutrino oscillations and thus enhance (or
suppress) the genuine signals of $CP$ or $T$ violation.
As for the long-baseline neutrino experiment under consideration,
the matter-induced effect in the $T$-violating asymmetry ${\cal A}_T$ is 
negligibly small. The matter effect on the $CP$-violating asymmetry
in vacuum cannot be neglected, but the former is unlikely to fake the
latter completely. We confirm numerically that the
relationship ${\cal A}_T = {\cal A}_{CP}$ in vacuum becomes violated
in matter.

If the earth-induced matter effects can well be controlled, it is 
possible to recast the fundamental flavor 
mixing matrix $V$ from a variety of measurements of 
neutrino oscillations. Such a goal is expected to be
reached in the neutrino factories \cite{Factory}.

In summary, we have derived an exact and compact formula to 
show the analytical relationship between the fundamental 
neutrino mixing matrix and the matter-corrected one within the
framework of three lepton families. This model-
and parametrization-independent result can be particularly 
useful for the study of flavor mixing and $CP$ violation in the
long-baseline neutrino experiments. An extension of 
the present work, in which the mixing of a sterile neutrino with
three active neutrinos can be incorporated, is in progress.

{\it Acknowledgment:} ~ The author would like to thank H. Fritzsch
for useful discussions.

\newpage

\begin{figure}[t]
\epsfig{file=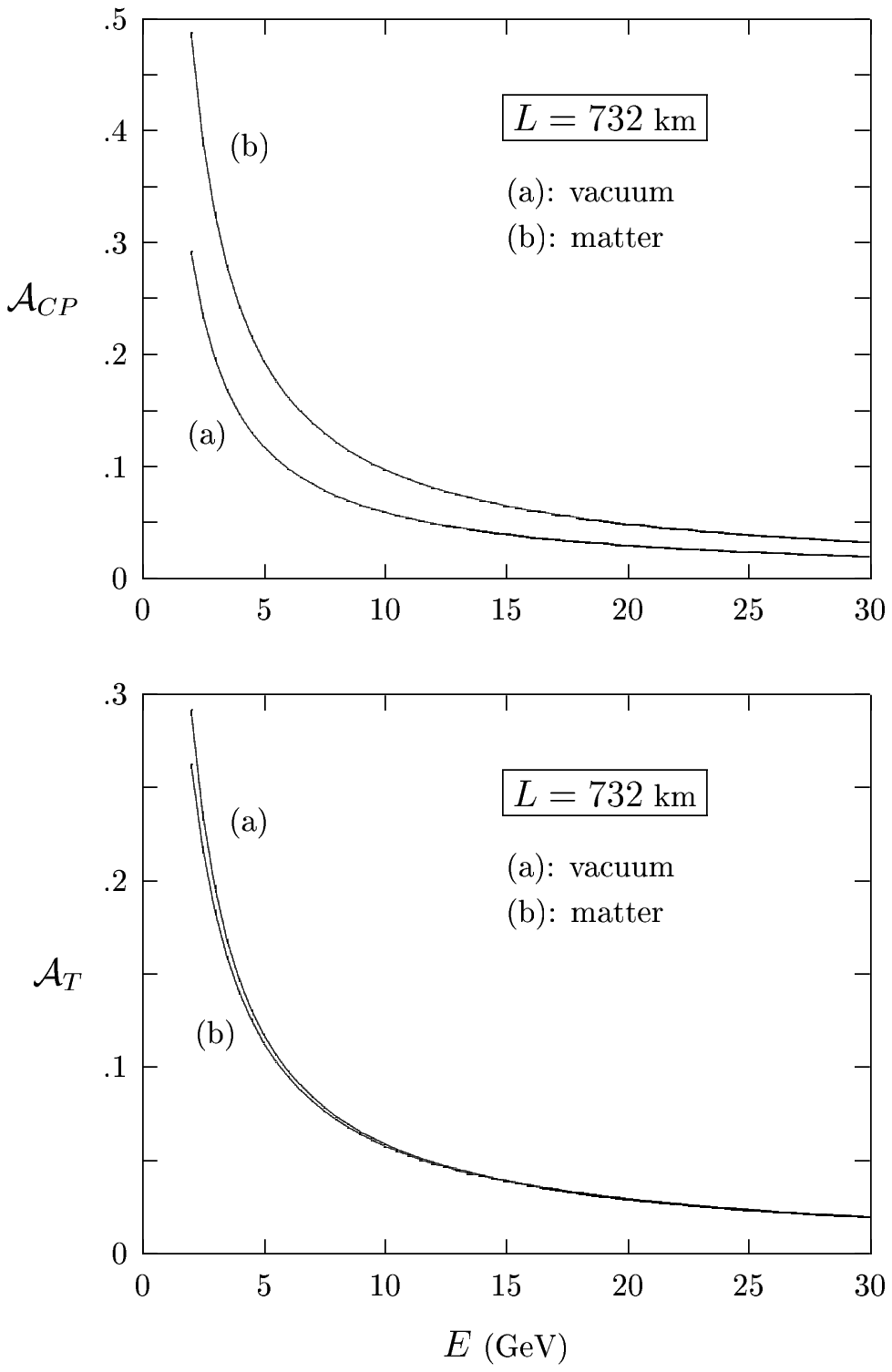,bbllx=1.5cm,bblly=-4cm,bburx=20cm,bbury=28cm,%
width=15.5cm,height=25cm,angle=0,clip=}
\vspace{-12.7cm}
\caption{Illustrative plots for
matter effects on ${\cal A}_{CP}$ (between $\nu_e \rightarrow \nu_\mu$
and $\bar{\nu}_e \rightarrow \bar{\nu}_\mu$ transitions) 
and ${\cal A}_T$ (between $\nu_e \rightarrow \nu_\mu$ and
$\nu_\mu \rightarrow \nu_e$ transitions) in a long-baseline
neutrino experiment with $L= 732$ km,
where $\Delta m^2_{21} = 5\cdot 10^{-5} ~ {\rm eV}^2$ and
$\Delta m^2_{31} = 3\cdot 10^{-3} ~ {\rm eV}^2$ have typically been input.}
\end{figure}
\begin{figure}[t]
\epsfig{file=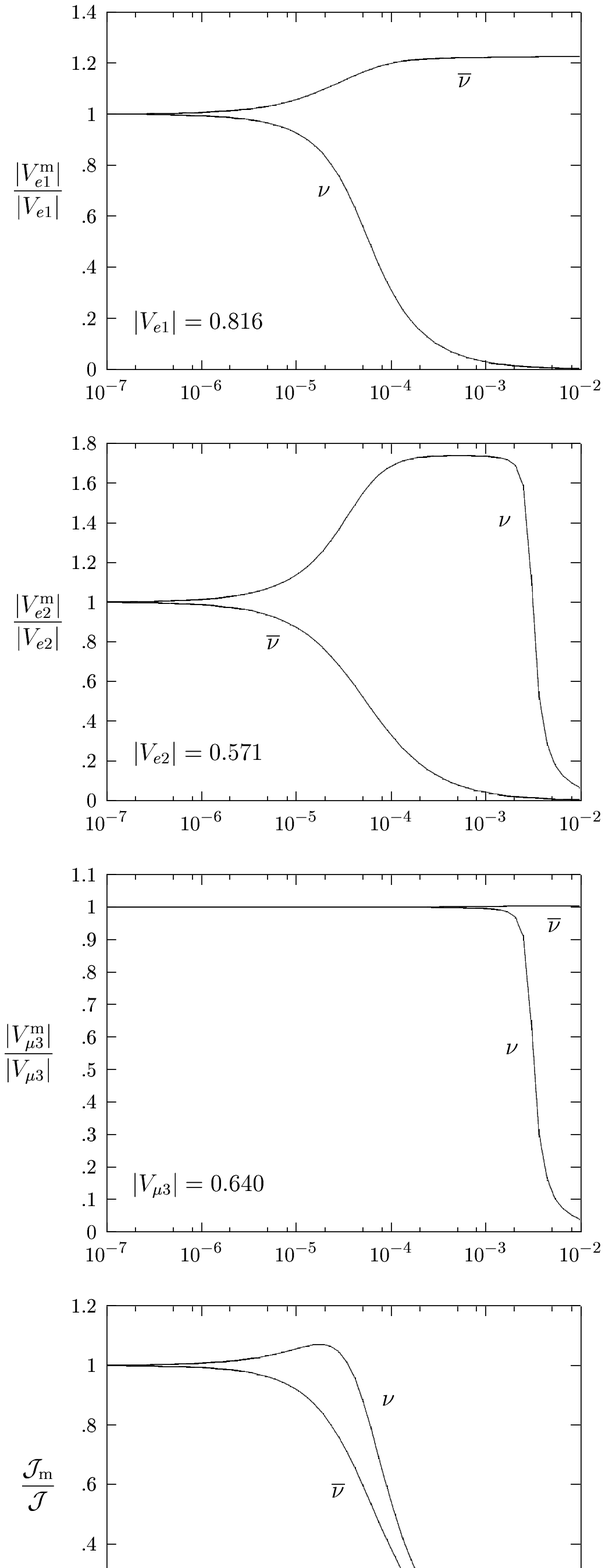,bbllx=1.6cm,bblly=-4cm,bburx=20cm,bbury=28cm,%
width=15.5cm,height=23cm,angle=0,clip=}
\vspace{-0.9cm}
\caption{Illustrative plots for 
matter effects on $|V_{e1}|$, $|V_{e2}|$, $|V_{\mu 3}|$ and $\cal J$
associated with neutrinos ($+A$) and antineutrinos ($-A$), 
where $\Delta m^2_{21} = 5\cdot 10^{-5} ~ {\rm eV}^2$ and
$\Delta m^2_{31} = 3\cdot 10^{-3} ~ {\rm eV}^2$ have typically been input.}
\end{figure}


\newpage
\begin{thebibliography}{99}
\bibitem{SK} Y. Fukuda {\it et al.}, Phys. Rev. Lett. {\bf 81}
(1998) 1562; {\it ibid.} {\bf 81} (1998) 4279.
http://www-sk.icrr.u-tokyo.ac.jp/dpc/sk/.

\bibitem{LSND} C. Athanassopoulos {\it et al.}, Phys. Rev. Lett. 
{\bf 75} (1995) 2650.

\bibitem{KARMEN} B. Zeitnitz, talk given at 
{\it Neutrino '98}, Takayama, Japan, June 1998.

\bibitem{Long} P. Fisher, B. Kayser, and K.S. McFarland,
Ann. Rev. Nucl. Part. Sci. {\bf 49} (1999) 481; and
references therein.

\bibitem{Factory} B. Autin {\it et al.}, CERN 99-02 (1999); 
D. Ayres {\it et al.}, physics/9911009; and references therein.

\bibitem{Wolfenstein} L. Wolfenstein, Phys. Rev. D {\bf 17} (1978) 2369.

\bibitem{Barger80} V. Barger {\it et al.}, 
Phys. Rev. D {\bf 22} (1980) 2718.

\bibitem{Langacker} P. Langacker, J.P. Leveille, and J. Sheiman,
Phys. Rev. D {\bf 27} (1983) 1228.

\bibitem{Smirnov} S.P. Mikheyev and A. Yu Smirnov, Yad. Fiz. 
(Sov. J. Nucl. Phys.) {\bf 42} (1985) 1441.

\bibitem{Zaglauer} H.W. Zaglauer and K.H. Schwarzer, Z. Phys. C
{\bf 40} (1988) 273;
P.I. Krastev and S.T. Petcov, Phys. Lett. B {\bf 205} (1988) 84.

\bibitem{Kuo} For an early review with extensive references, see:
T.K. Kuo and J. Pantaleone, Rev. Mod. Phys. {\bf 61} (1989) 937.

\bibitem{Barger99} V. Barger {\it et al.}, hep-ph/9911524;
M. Freund {\it et al.}, hep-ph/9912457;
A. Cervera {\it et al.}, hep-ph/0002108.

\bibitem{Jarlskog} C. Jarlskog, Phys. Rev. Lett. {\bf 55} (1985) 1039.

\bibitem{FX99} H. Fritzsch and Z.Z. Xing, 
``Mass and Flavor Mixing Schemes of Quarks and Leptons'',
hep-ph/9912358, invited review article to appear in Prog. Part.
Nucl. Phys..

\bibitem{Harrison} P.F. Harrison and W.G. Scott, 
Phys. Lett. B {\bf 476} (2000) 349.

\bibitem{PDG98} C. Caso {\it et al.}, Eur. Phys. J. C {\bf 3} (1998).

\bibitem{CHOOZ} M. Apollonio {\it et al.}, Phys. Lett. B {\bf 420} (1998) 397.

\bibitem{Bahcall} J.N. Bahcall, P.I. Krastev, and A.Y. Smirnov,
Phys. Rev. D {\bf 58} (1998) 096016 (1998); 
M.C. Gonzalez-Garcia {\it et al.}, hep-ph/9906469;
G.L. Fogli {\it et al.}, hep-ph/9912231.

\end{thebibliography}
\end{document}